\documentclass{emulateapj}
\usepackage{graphicx}
\begin{document}

\def\etal{et al.\ \rm}

\title{Effects of Gravitational Lensing in the Double Pulsar System J0737-3039}

\author{Dong Lai\altaffilmark{1} and Roman R. Rafikov\altaffilmark{2}}
\altaffiltext{1}{Department of Astronomy, Cornell University, 
Ithaca, NY 14853; dong@astro.cornell.edu}
\altaffiltext{2}{IAS, Einstein Dr., Princeton, NJ 08540; rrr@ias.edu}


\begin{abstract}
We investigate the effects of gravitational lensing in the binary 
pulsar system J0737-3039. Current measurement of the orbital inclination 
allows the millisecond pulsar (A) to pass very close
(at $R_{\rm min}\approx 4000$~km) in projection to the companion 
pulsar (B), with $R_{\rm min}$ comparable to the Einstein 
radius ($\approx 2600$~km). 
For this separation at the conjunction, lensing causes small
(about $10\%$) magnification of the pulsar A signal on 
a timescale of several seconds, and displaces the pulsar image on 
the sky plane by about $1200$~km. More importantly, lensing 
introduces a correction (of several $\mu s$) to the conventional 
Shapiro delay formula used in pulsar timing analysis, and gives 
rise to a geometric time delay together with the delays associated
with the pulsar spin period.
These lensing effects can influence the determination 
of the system parameters by both timing and scintillation studies.
Given the current uncertainty in the orbital inclination, 
more extreme manifestations of lensing (e.g. magnification by a factor 
of several) are possible. We compare our predictions with the existing 
observations and discuss the possibility of detecting gravitational lensing 
signatures in the system. The anomalously high point in A's lightcurve close
to superior conjunction might be caused by gravitational lensing.
\end{abstract}

\keywords{pulsars: general ---  stars: neutron --- pulsars:
individual (J0737-3039A, J0737-3039B) --- gravitational lensing
--- binaries: general}


\section{Introduction}
\label{sect:intro}


The binary pulsar system J0737-3039 consists of a 
millisecond pulsar (pulsar A with a period  $P_A=23$ ms) and 
a normal pulsar (pulsar B with $P_B=2.8$ s) in a 
$2.4$ hr orbit with eccentricity $e=0.088$ (Burgay \etal 2003). 
The system has a nearly edge-on orientation with respect to 
our line of sight, leading to a brief (about $27$~s) eclipse of 
pulsar A by the magnetosphere of pulsar B during each orbit
at A's superior conjunction (Lyne \etal 2004; Kaspi 
\etal 2004; McLaughlin \etal 2004). The depth of the eclipse 
is quite significant, corresponding to an absorption optical 
depth of order a few (Kaspi \etal 2004). The eclipse is asymmetric: 
it is deeper and longer after the conjunction, indicating that 
the optical depth increases from the eclipse ingress to egress. 

Timing observation (including measurement of Shapiro delay) 
provided the first determination of the system's inclination angle
with respect to our line of sight, $i=87^\circ\pm 3^\circ$ (Lyne \etal 2004).
Recently, Coles \etal (2004) have used the correlation of 
the interstellar scintillations of both pulsars to obtain a
remarkable constraint on the minimum distance between A and B on the 
plane of the sky, $R_{\rm min}=4000\pm 2000$~km, 
corresponding to $i=90.29^\circ\pm 0.14^\circ$. 
As we show in this {\it letter}, such a small projected separation of A and B
gives rise to interesting gravitational lensing effects during each
orbit. 


\section{Gravitational Lensing in J0737}
\label{sect:cyclo}


At the superior conjunction of pulsar A, the distance between
A and B projected to our line of sight 
is $a_\parallel=a|\sin i|(1-e^2)/(1+e\sin\omega)$, where
$a=8.79\times 10^5$~km is the semimajor axis of the orbit, 
$\omega$ is the longitude of periastron 
(at present $\omega=74^\circ$, and $\dot\omega =16.9^\circ/$yr).
With pulsar B mass $M_B=1.25~M_\odot$, the Einstein radius is
\begin{eqnarray}   
R_E=(2R_g a_\parallel)^{1/2}\simeq(2R_g a)^{1/2} =2550~{\rm km},
\label{eq:ein_rad}
\end{eqnarray}
where $R_g=2GM_B/c^2=3.69$~km.
This is 
already 
comparable to the minimum projected A-B distance 
$R_{\rm min}=a_\parallel |\tan i|^{-1}\simeq 4000\pm 2000$~km. 
Given the present uncertainty of the measurement, the real $R_{\rm min}$
might be smaller than $R_E$, increasing the strength of 
gravitational lensing.

The typical timescale of lensing is 
$t_0\sim R_E/v\approx 4$~s (where $v\approx 680$~km s$^{-1}$ is the 
relative velocity between the two pulsars), considerably 
shorter than the total duration ($\sim 27$~s) of the magnetospheric eclipse.
Gravitational lensing causes variation of the 
pulse flux during the eclipse, with the magnification factor (Paczy\'nski 1996)
\begin{eqnarray}
A=\frac{u^2+2}{u\sqrt{u^2+4}},~~~u=\frac{R_s}{R_E}.
\label{eq:magnif}
\end{eqnarray}
Here $R_s=r(1-\sin^2i\sin^2\psi)^{1/2}$ is the separation 
between the two pulsars in the plane of the sky, with
$r=a(1-e^2)/(1+e\cos\phi)$ and $\psi=\phi+\omega$
being the physical separation between A and B and the longitude 
from ascending node, respectively ($\phi$ is a true anomaly 
measured from periastron).
Let $\psi=\pi/2+\Delta\psi$ (the A-B conjunction corresponds to $\Delta\psi=0$)
and $i=\pi/2+\Delta i$, we have $R_s\simeq a_\parallel
\left[(\Delta i)^2+(\Delta\psi)^2\right]^{1/2}$ for $|\Delta i|,
|\Delta\psi|\ll 1$.

For the current best estimate of $R_{\rm min}\simeq a_\parallel \Delta i\simeq 
4000$~km ($i=90.29^\circ$) one obtains $A_{\rm max}=1.13$. 
Such $10\%$ flux variation due to lensing 
could have been marginally detectable with currently available 
observations (Kaspi \etal 2004) at $820$ MHz if the superior 
conjunction of pulsar A were not accompanied by its
simultaneous eclipse by the pulsar B's magnetosphere
(see \S \ref{subsect:detection}). 
This may however be possible when more data on the eclipse become available. 

Smaller $R_{\rm min}$ leads to stronger magnifications: 
for $R_{\rm min}=2000$~km ($i=90.14^\circ$), 
corresponding to $1\sigma$ deviation from the mean,
we find $A_{\rm max}=1.6$; for $R_{\rm min}=1000$~km
($i=90.07^\circ$), $A_{\rm max}=2.8$. 
Thus, the amplification of pulsar A signal by a factor 
of several is not inconsistent with the currently measured
inclination of the system --- in the end even the purely
edge-on orientation of the system ($i=90^\circ$) deviates
only by $2\sigma$ from the measurement of Coles \etal (2004).


\subsection{Time Delay}
\label{subsect:delay}

\begin{figure}
\plotone{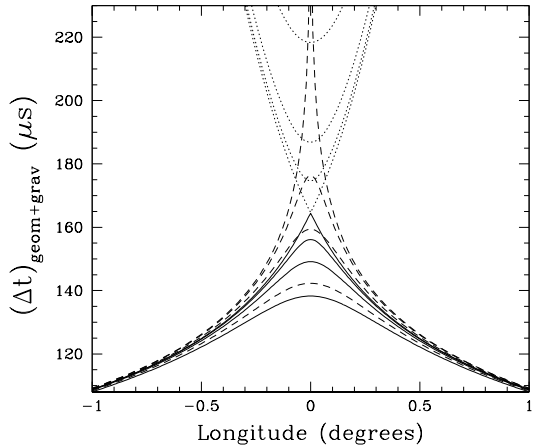}
\vskip -0.8cm 
\caption{
The combined geometric delay and gravitational
(Shapiro) delay of pulsar A's signal (for the ``+'' image)
including the lensing 
effect by pulsar B (solid lines). Timing perturbations 
associated with the pulsar spin period (see the text) are not 
included. The dashed lines show the Shapiro delay when  
lensing is not included. Both the solid and dashed lines 
correspond to $R_{\rm min}=0,~1000,~2000$ and $4000$~km
from top to bottom. The dotted lines show the time delay associated
with the ``$-$'' images (with $R_{\rm min}$ increasing from bottom to top).
Note that the ``$-$'' images are usually much fainter than the 
``+'' images. The longitude is measured from
the superior conjunction of pulsar A (i.e., when A is behind B). 
\label{fig:delay}}
\end{figure}

Gravitational lensing has important effect on the delay
of the pulse arrival time (Schneider 1990). The geometric 
time delay is given by
\begin{eqnarray}
(\Delta t)_{\rm geom}=\frac{R_g}{c}\left(\frac{\Delta R_\pm}
{R_E}\right)^2,
\label{eq:delay}
\end{eqnarray}
where $\Delta R_\pm=|R_\pm-R_s|$ is the relative distance in the plane 
of the sky between the image position of the source,
$R_\pm=0.5(R_s\pm \sqrt{R_s^2+4R_E^2})$, and its fiducial position 
in flat space, $R_s$; ``$+ (-)$'' refers to the image which 
is closer (further) from the source. The magnification factor
of the image is $A_\pm=(A\pm 1)/2$. For $u=R_s/R_E\to 0$ one finds
$\Delta R_\pm\to R_E$ and 
the two images merge into an Einstein  ring;
for $u\gg 1$ one has $\Delta R_+\to R_E u^{-1}$ and
$\Delta R_-\to R_E u$, but the ``$-$'' image contributes negligible light
(Paczynski 1996). For J0737, the maximum possible geometric delay is
$(\Delta t)_{\rm geom}^{\rm (max)}=R_g/c=12~\mu s$,
achieved for $u=0$ ($i=90^\circ$ and $\psi=90^\circ$).

Gravitational (Shapiro) delay amounts to as much as $100~\mu s$ 
in J0737. If lensing is neglected, the Shapiro delay is 
given by the standard formula (Blandford \& Teukolsky 1976)
\begin{eqnarray}
(\Delta t)_{\rm grav}^{\rm (no~lens)}={R_g\over c}
\ln \left({1+e\cos\phi\over 1-\sin i\sin\psi}\right).
\label{eq:Shap_no_lens}
\end{eqnarray}
Note that for $i=90^\circ$, this expression diverges at $\psi=90^\circ$ 
since in the absence of lensing the ray would pass through the 
infinitely deep potential of pulsar B. 
This is an artifact of neglecting the light bending caused by 
the lensing --- the correct formula including the lensing effect is
\footnote{The formula given by Schneider (1990) is incorrect; e.g.,
it does not reduce to the standard formula in the no-lensing limit.}
\begin{eqnarray}
(\Delta t)_{\rm grav}=-{R_g\over c}
\ln \left[{\sqrt{r_\parallel^2+R_\pm^2}-r_\parallel\over a(1-e^2)}
\right],
\label{eq:Shap_lens}
\end{eqnarray}
where $r_\parallel=r\sin i\sin \psi$ is the separation of A and 
B along the line of sight. It is easy to see that for $R_E=0$ (no lensing), 
this expression reduces to the usual (no lens) result. Unlike
eq.~(\ref{eq:Shap_no_lens}), the lensing-corrected expression
(\ref{eq:Shap_lens}) never diverges: the maximum possible delay is
$(\Delta t)_{\rm grav}^{\rm (max)}=(R_g/c)\ln[a(1-e^2)/R_g]
=152~\mu s$, again achieved for $i=\psi=90^\circ$.

Finally, light bending causes radio pulses 
detected on Earth to be emitted in a direction slightly 
different from our line of sight, by $\Delta R_+/r_\parallel$ for the
``$+$'' image. This corresponds to a different spin phase of the pulsar at 
the moment of pulse emission than the phase would be in the absence of 
lensing, leading to another timing perturbation proportional to the pulsar 
spin period\footnote{There is also additional contribution to the delay 
due to the relativistic aberration of light (Smarr \& Blandford 1976).} 
(Schneider 1990). Near the 
superior conjunction of pulsar A this delay amounts to several $\mu s$ for 
the pulsar A signal\footnote{The corresponding delay of the signal of pulsar B 
at its superior conjunction is $\sim 10^{-3}s$ (because of the large $P_B$).}.
This type of delay is very interesting since it is sensitive to the 
orientation of the pulsar spin axis (Doroshenko \& Kopeikin 1995). 
Its contribution is not symmetric with respect to the conjunction point 
which makes possible its separation from other timing contributions. 
Because of the importance of this delay for 
constraining the system's geometry we postpone its detailed study
to a separate paper (Rafikov \& Lai 2005).

In Figure \ref{fig:delay} 
we compare the time delay with and without the geometric delay and 
corrections to the Shapiro delay caused by lensing effects in 
J0737 for the dominant, ``$+$'' image near the point of eclipse of pulsar A
for different values of $R_{\rm min}$. We do not include the delays 
associated with the pulsar spin period in this Figure (this is equivalent to
setting pulsar spin period to zero). One can see that for 
$R_{\rm min}\le 4000$~km,
$(\Delta t)_{\rm geom}+(\Delta t)_{\rm grav}$ differ from the conventional
Shapiro delay formula by $\gtrsim 4~\mu s$. This emphasizes the importance of 
accounting for the lensing effects which affect the estimate of the system's 
parameters. Both the geometric delay
and the correction to the Shapiro delay are localized near the 
conjunction of the pulsars and their variation has a typical timescale 
of $t_0$ (unlike the Shapiro delay 
without lensing, which varies on the scale of the orbital 
period of the system).

For $R_s\sim R_E$, assuming that pulsar A
is a point source (see \S\ref{subsect:size}),
one should expect to see a splitting of the 
original pulsar A pulse profile in two because of the two 
gravitationally lensed images, so that the modified pulse profile 
is a sum of the two weighted replicas of the normal pulsar A pulse 
profile slightly shifted in time with respect to each other. 
The relative time delay 
between the two contributions and their amplitudes should vary 
in the course of the eclipse as $R_s$ varies. This
modification of the pulse profile may be difficult to 
detect: for $R_{\rm min}=4000$~km, one finds that at the conjunction
$\Delta t=(\Delta t)_{\rm geom}+(\Delta t)_{\rm grav}
\simeq 138~\mu s$ and $218~\mu s$ for the two images, with 
the image magnification factor $A_+=1.07$ and $A_-=0.07$.
Thus, the contribution to the total pulse 
profile coming from the ``$-$'' image with longer time delay 
(see Fig.~1) is very small. Also note that the ``$-$'' image
samples the deeper interior of the pulsar B magnetosphere than the 
``$+$'' image does, which causes its stronger absorption. 
Nevertheless, the two-image time delay corrections described above 
can in principle be detected with exquisite timing observations, especially
if $R_{\rm min}$ is small (e.g., for  
$R_{\rm min}=1000$~km, the time delays of the two images are
$\Delta t\simeq 156~\mu s$ and $175~\mu s$, with 
$A_+=1.9$ and $A_-=0.9$; see Rafikov \& Lai 2005 for more details).


\subsection{Has Lensing Been Detected in J0737?}
\label{subsect:detection}


\begin{figure}
\plotone{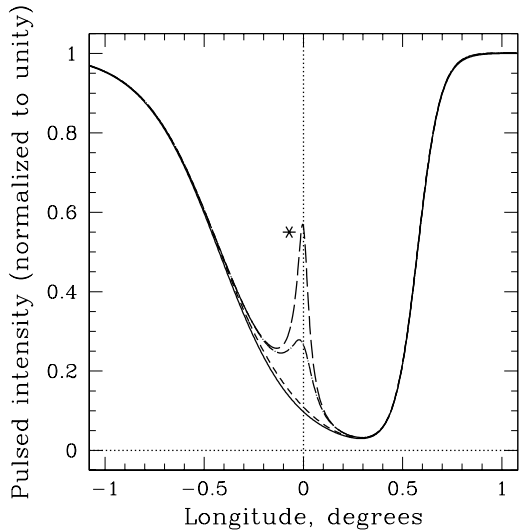}
\caption{
Model lightcurve of the eclipse with the effect of the gravitational 
lensing folded in. Modulation of the lightcurve by pulsar B's rotation is
neglected (see text). The solid curve represents unmagnified eclipse
profile. Other curves correspond to $R_{\rm min}=4000$~km 
({\it short-dashed}), $1000$~km ({\it dot-dashed}) and 
$450$~km ({\it long-dashed}) 
(we assume that the two images suffer the same amount of attenuation; 
see text). The star approximately marks the position of the flux anomaly
during one of the eclipses reported by Kaspi \etal (2004).
\label{fig:lightcurve}}
\end{figure}

Detecting any signature of lensing is made difficult in the
case of J0737 by the strong flux attenuation of pulsar A
at its superior conjunction. Magnetospheric eclipse affects both 
the magnification measurements and (to a lesser extent) timing measurements.

Because of the short lensing timescale compared to the 
eclipse duration, the lensing magnification should 
manifest itself as a spike in the lightcurve of pulsar A 
in the inner part of the eclipse profile. 
Using the published parameters of the eclipse (Kaspi \etal 2004)
we model pulsar A lightcurve in Figure \ref{fig:lightcurve} at $1400$~MHz
for different values of $R_{\rm min}$\footnote{Note that
the lightcurve of Kaspi \etal was based on averaging over 2-second
time intervals of the pulses. On a finer time resolution ($~0.3$~s),
the lightcurve shows many spikes and dips due to 
the modulation of A's signal by B's rotating magnetosphere
(McLaughlin et al.~2004).}. The shape of the spike is skewed 
because the magnification profile is convolved with the eclipse 
lightcurve which exhibits an increase in optical depth with time. This
causes the spike to occur slightly {\it before} the superior 
conjunction of pulsar A.

It is rather curious that strong fluctuations in the flux 
of pulsar A during the eclipse have already been detected.
Kaspi \etal (2004) report a detection of a $5\sigma$ outlier
in the eclipse lightcurve at $1400$ MHz just seconds before
the conjunction (see their Figure 2). The same 
phenomenon is also seen in the measurements by McLaughlin \etal 
(2004) who demonstrate the variation of the eclipse profile
with the rotation phase of pulsar B. In their lightcurves,
among the multiple spikes and dips caused by B's magnetosphere,
an anomalously high flux point has again been seen prior to 
the conjunction (see their Figures 1 and 3). Gravitational lensing 
can explain the outlying point in the Kaspi \etal measurements if 
$R_{\rm min}\simeq 400$~km ($i=90.029^\circ$),
corresponding to the magnification of $A_{\rm max}=6$. 

The lensing interpretation of the observed 
flux anomalies during the eclipse is complicated by the fact that 
spikes are not observed during all eclipses and there is
only one strongly outlying point. 
Also, in the rotation phase-resolved measurements of 
McLaughlin \etal (2004), 
one should expect to see a flux enhancement of A
at the B pulse phase $0.25$
similar to that seen at the phase $0.75$, since in both phases
the unmagnified fluxes are quite significant.
The error bars on all these measurements are currently 
too large for us to decide whether the flux anomalies are 
truly caused by gravitational lensing.

Current timing measurements of pulsar A have not been able to 
detect time delay during the eclipse, only the upper
limit on the change in the dispersion measure during the
eclipse has been set at the level of $D<0.016$~pc~cm$^{-3}$ 
(Kaspi \etal 2004). This corresponds to a current timing 
resolution of about $10-20~\mu s$ which makes the geometric delay
and lensing correction to the Shapiro delay 
hard to observe at present --- the effects are 
of order 
$4~\mu s$ for $R_{\rm min}=4000$ km (see Fig.~\ref{fig:delay}).
However, with the improved timing resolution at the level of $1~\mu s$, 
one would be able to detect them even if 
$R_{\rm min}$ is as large as $10^4$ km (at this $R_{\rm min}$ magnification 
is too small to be detectable). Timing can be a much better probe of 
lensing than the magnification if $R_{\rm min}\gtrsim R_E$. 


\subsection{Size of the Emission Region of Pulsar A}
\label{subsect:size}


Under many circumstances lensing can be used as a probe of the size 
of the source (Paczy\'nski 1996). The size of the emitting region in 
pulsar A is limited from above by the light cylinder radius,
$R_{\rm em,A}< R_{L,A}=c/\Omega_A\approx 1000$~km. 
Finite source size limits the magnification to always stay below 
$\sqrt{4(R_E/R_{\rm em,A})^2+1}\approx 5(R_{L,A}/R_{\rm em,A})$
(e.g., Witt \& Mao 1994), and any observed deviation from the 
point-source amplification formula could constrain $R_{\rm em,A}$.
But unless $R_{\rm min}\lesssim R_{\rm em,A}<1000$~km this limit 
cannot be achieved and the finite source size effects are small. 
If the emitting region size in the millisecond pulsar 
is much smaller than $R_{L,A}$, the detection of the source 
size through lensing would be unlikely.


\subsection{Effects of Orbital Evolution}
\label{subsect:evolution}


Effects of general relativity are very strong in the J0737
system. They drive the evolution of its orbital
parameters and affect its geometry, which directly translates
into the change of the lensing signal. Geodetic precession 
causes a wobble of the orbital plane of the system on a period 
of $75$ yrs (Burgay \etal 2003). Unfortunately, the amplitude of 
the inclination variation caused by this is very small, of order
$(S_A/L_{\rm orb})\sin\theta_{SL}\sim 4\times 10^{-5}\sin\theta_{SL}$, 
where $S_A$ is the spin angular momentum of pulsar A, 
$L_{\rm orb}$ is the orbital angular momentum,
and $\theta_{SL}$ is the angle between the spin axis and the orbital
angular momentum axis.
Thus, geodetic precession is not going to affect gravitational 
lensing in J0737 unless $|\Delta i| \lesssim 0.02^\circ$.

More important would be the effect of the periastron advance
on a timescale of $21$ yrs. It causes the projected distance  
between the two pulsars at their conjunction,
$R_{\rm min}$,
to vary from $a (1-e) |\cos i|$ to $a(1+e)|\cos i|$. 
Thus, in the course of the apsidal precession, $R_{\rm min}$
varies by about $18\%$. This causes a change in the lensing 
signatures (both magnification and timing)
at a similar level of $\sim 10\%$ and might be detectable in the 
future. 


\section{Discussion}
\label{sect:disc}

Additional propagation effects are caused by
the plasma refractivity in the magnetosphere of pulsar B 
during the eclipse (e.g. Emmering \& London 1990). 
Assuming that the magnetospheric charge density is $f$ times
the Goldreich-Julian density (Goldreich \& Julian 1969),
the deviation of the index of refraction from unity is of
order $\Delta n\sim 10^{-7}f/\nu^2$, where $\nu$ is the radio wave frequency 
in GHz and density is evaluated at the Einstein radius $R_E$
\footnote{This estimate assumes cold plasma in the
magnetosphere. For a relativistic plasma, $\Delta n$ is reduced further,
making the refractive effect of the plasma even less important.}.
By contrast, the effective index of refraction due to lensing 
deviates from unity by $R_g/R_E=1.5\times 10^{-3}$.
For the J0737 system, $f$ is likely of order $100$ 
(Rafikov \& Goldreich 2005), in which case the propagation effects 
due to the plasma in the vicinity of pulsar B 
are not important compared to the gravitational lensing effect.

Since the gravitational lensing signal is strongly dependent upon
the minimum projected separation $R_{\rm min}$ of the pulsars in 
the plane of the sky, the detection of lensing effects could provide 
tighter constraints on the system's inclination,
which is currently known at an accuracy of $0.14^\circ$. 
Timing would be a good probe of the inclination if $R_{\rm min}\gtrsim R_E$ 
while the magnification signal would also be useful for handling the case 
$R_{\rm min}\lesssim R_E$.

We have already mentioned that the correction to the Shapiro 
delay and the additional geometric delay caused by lensing at the binary 
conjunction may affect the estimation of the system's parameters.
This suggests that lensing must be accounted for 
when dealing with the effects of general relativity in J0737.
Gravitational light bending is also relevant for the measurements using the 
correlation of scintillations of the two pulsars. The
analysis of Coles \etal (2004) assumes that 
pulsar A's trajectory in the plane of the sky is
a straight line at the conjunction. This is not true 
if lensing is strong because it displaces the  
position of pulsar A on the sky along a {\it curved} path 
deviating from the straight line trajectory by $\Delta R_+$. 
Even for $R_{\rm min}=4000$~km, the shift can be as large as $1200$~km,
almost $30\%$ of the minimum separation claimed by Coles \etal (2004). 
This emphasizes the importance of incorporating 
lensing effects into the analysis of scintillation 
correlations which could further refine the
estimate of inclination obtained by Coles \etal (2004). 

Light bending also causes the radio beam of pulsar A to pass through
the regions in the pulsar B magnetosphere which would otherwise not be 
sampled by the beam. Since the magnetospheric absorption depends strongly 
on the distance from pulsar B (Rafikov \& Goldreich 2005), 
one expects gravitational lensing to have an important effect on the 
eclipse profile of pulsar A. Variations in the pattern of magnetospheric 
absorption would then be expected on a timescale of apsidal precession 
because of the change in the minimum projected distance between the 
two pulsars at conjunction (see \S \ref{subsect:evolution}). 

Pulsar A can produce gravitational lensing of pulsar B in exactly the same way
as B does it with pulsar A. The Einstein radius of pulsar A is
slightly larger than that of pulsar B, $2640$~km instead of $2550$~km,
and the measurements would probably be cleaner in this case
(although the timing accuracy of B is not as good as that of A)
since the radio beam of pulsar B would not pass through the magnetosphere 
of pulsar A (with light cylinder radius $R_{L,A}\approx 1000$~km).
Unfortunately, pulsar B is currently very weak
at the orbital phases when it is behind pulsar A (Ramachandran \etal 2004) 
which makes it hard to detect lensing 
in this orbital configuration. However, the system's geometry will 
change due to the effects of general relativity and  
pulsar B might become active and bright at its superior conjunction
(Jenet \& Ransom 2004). This may give us additional handle on the 
gravitational lensing in J0737 and permit the determination of 
the pulsar B spin axis orientation (see \S \ref{subsect:delay}).

\acknowledgements 
We are grateful to Peter Goldreich for stimulating
discussions and Maura McLaughlin for useful comments.
DL thanks the Institute for Advanced Study 
for hospitality where this work was carried out; he is
supported in part by NSF grant AST 0307252 and NASA grants
NAG 5-12034 and SAO-TM4-5002X. 
RRR thankfully acknowledges the financial support 
by the W. M. Keck Foundation and NSF via grant PHY-0070928. 


\end{document}